\documentclass[doublecol]{epl2} 
\usepackage{amsmath}
\renewcommand{\revision}[1]{#1}

\title{Andreev tunneling in charge pumping with SINIS turnstiles}

\author{T. Aref\inst{1,2} \and V. F. Maisi\inst{2} \and M. Gustafsson\inst{3} \and P. Delsing\inst{3} \and J. P. Pekola\inst{1}}
\shortauthor{T. Aref \etal}

\institute{                    
  \inst{1} Low Temperature Laboratory, Aalto University - P.O. Box 13500, 00076 Aalto, Finland\\
  \inst{2} Centre for Metrology and Accreditation (MIKES) - P.O. Box 9, 02151 Espoo, Finland\\
  \inst{3} Department of Microtechnology and Nanoscience (MC2), Chalmers University of Technology, SE-412 96 G\"oteborg, Sweden
}
\abstract{We present measurements on hybrid \revision{single-electron} turnstiles with superconducting leads contacting a normal island (SINIS). We observe Andreev tunneling of electrons influencing the current plateau characteristics of the turnstiles under radio-frequency pumping. The data is well accounted for by numerical simulations. We verify the dependence of the Andreev tunneling rate on the turnstile's charging energy. Increasing the charging energy effectively suppresses the Andreev current.}

\begin{document}

\maketitle

\section{Introduction}
At the present time, there does not exist a quantum current standard though such standards exist for both voltage and resistance. A strong candidate for such a standard is a turnstile with a small normal metal island connected by tunnel junctions to superconducting leads (SINIS)\cite{Pekola2008}. These hybrid turnstiles pump electrons one at a time producing a well-defined current of $I=ef$ where $e$ is the electron charge and $f$ is the frequency of pumping. Understanding and eliminating error processes in these turnstiles is vital for realizing a quantum metrological triangle (QMT), a key goal in metrology \cite{Flowers2004}. In closing the QMT, the three standards of voltage, resistance and current would be compared against each other via Ohm's law. Closing the QMT will allow the most accurate comparison of \revision{the Josephson constant} $K_J=2e/h$ and \revision{the von Klitzing constant} $R_K=h/e^2$ (i.e. the charge of the electron and Planck's constant) to date.

\revision{There exist several potential quantum current standard candidates. The NIST seven junction pump has demonstrated a current accuracy of $1.5$ parts in $10^{8}$ \cite{Keller1996, Keller1999} but is limited to maximum currents of approximately 1 pA. Other candidates with the potential for metrological accuracy at metrologically relevant currents include semiconductor tunable barrier pumps, charge-coupled device pumps\cite{Fujiwara2004}, tunnel junction pumps\cite{Geerligs1990, Pothier1992, Lotkhov2001}, quantum dot pumps\cite{Blumenthal2007}, surface acoustic wave pumps\cite{Shilton1996} and quantum phase slip nanowire pumps\cite{Mooij2006}. In general, the accuracy of these quantum currents standards is not yet comparable to the accuracy of the quantum Hall and Josephson effect standards. A semiconducting quantum dot parallel pump recently demonstrated an accuracy of 1.5 in $10^{5}$ at 54 pA \cite{Giblin2010}. The minimum required current magnitude for closing the metrological triangle is about 100 pA with an accuracy of about 1 part in $10^{8}$, which has not yet been achieved.}

The SINIS turnstile operates by using energy barriers to control the flow of electrons. The normal metal island is capacitively coupled to a gate electrode. By applying an appropriate voltage to  the gate, single electrons can be added and removed from the normal metal island. In essence, the SINIS turnstile operates as a \revision{single-electron} \revision{transistor,} with the superconducting gap providing extra protection against unwanted tunneling of the electrons. 

There are multiple transport processes that can occur in a  SINIS turnstile. The dominant one employed in charge pumping is sequential single electron tunneling through the insulating barrier. The dominant \revision{two-electron} process causing errors is Andreev tunneling. In Andreev tunneling, an electron in the normal metal is reflected as a hole (or a hole as an electron) at the interface of the NIS junction.  This forms (or removes) a Cooper pair in the superconducting electrode and can be alternatively viewed as two electrons tunneling simultaneously across the insulating barrier\cite{Hergenrother1994}. It has been shown that errors arising from sequential $1e$-tunneling in turnstiles such as environmental activation can be effectively suppressed by proper filtering including an on-chip capacitively coupled ground plane\cite{PekolaMaisi2010}. Higher order, \revision{multiple-electron} error processes not eliminated by this filtering may then be observed.  Many experimental observations of Andreev currents have been reported previously\cite{Andreev1964, Blonder1982, Tinkham1996}. Andreev processes in devices with Coulomb blockade have been studied \revision{previously,} both theoretically \cite{Hekking1993} and experimentally \cite{Eiles1993, Hergenrother1994}. Only recently were individual Andreev reflection events detected in NIS junctions\cite{Maisi2011, Saira2011}.
 
  Here we demonstrate that the Andreev tunneling process is detectable in \revision{single-electron turnstile} current pumping plateaus and that the resulting error effect can be minimized by increasing the charging energy of the turnstile. The data is well accounted for by numerical simulations with reasonable parameter values and the observed behavior is physically intuitive. 

\begin{figure}[h!]
\onefigure{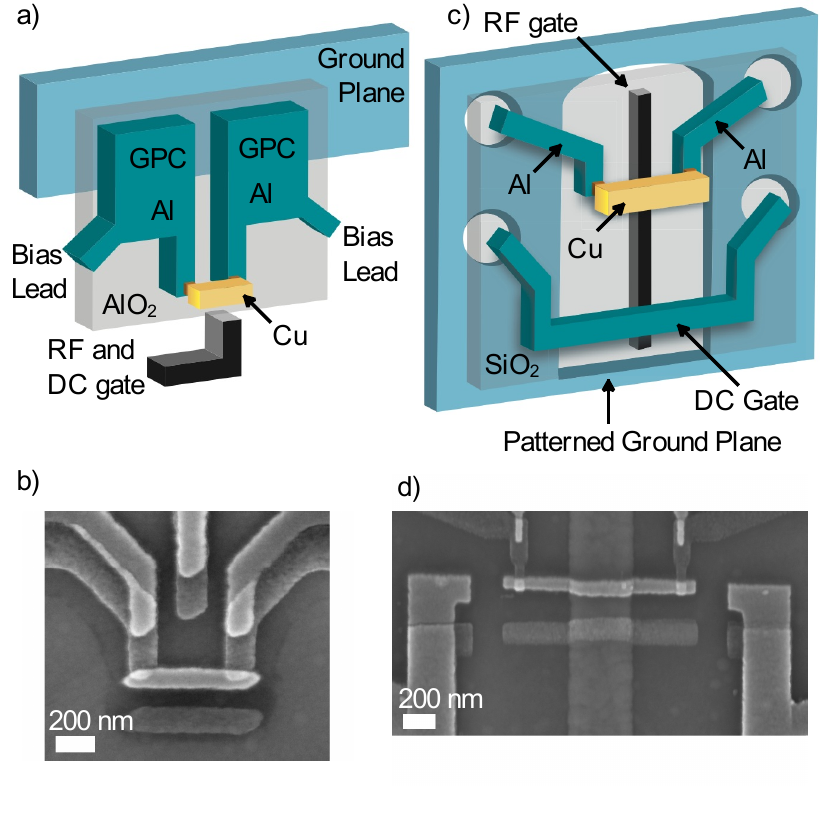}
\caption{\revision{Experimental set-up (a) Layout for sample of type one. The leads are coupled to the ground plane by large ground plane couplers (GPC) yet the turnstile itself is off the ground plane. (b) A SEM micrograph of the first sample type showing the turnstile and combined RF gate. The GPC and ground plane are not shown. This is sample S1 with junction area 50 nm by 125 nm. (c) The second sample set-up with the samples in the patterned part of the ground plane. The device design is modified to include a separate RF and DC gate.  (d) A SEM micrograph of the second sample type showing the turnstile and the DC gate on top of the RF gate. This is sample S4 with junction area 60 nm by 70 nm. }
}
\label{fig:setup}
\end{figure}

\begin{figure}
\onefigure{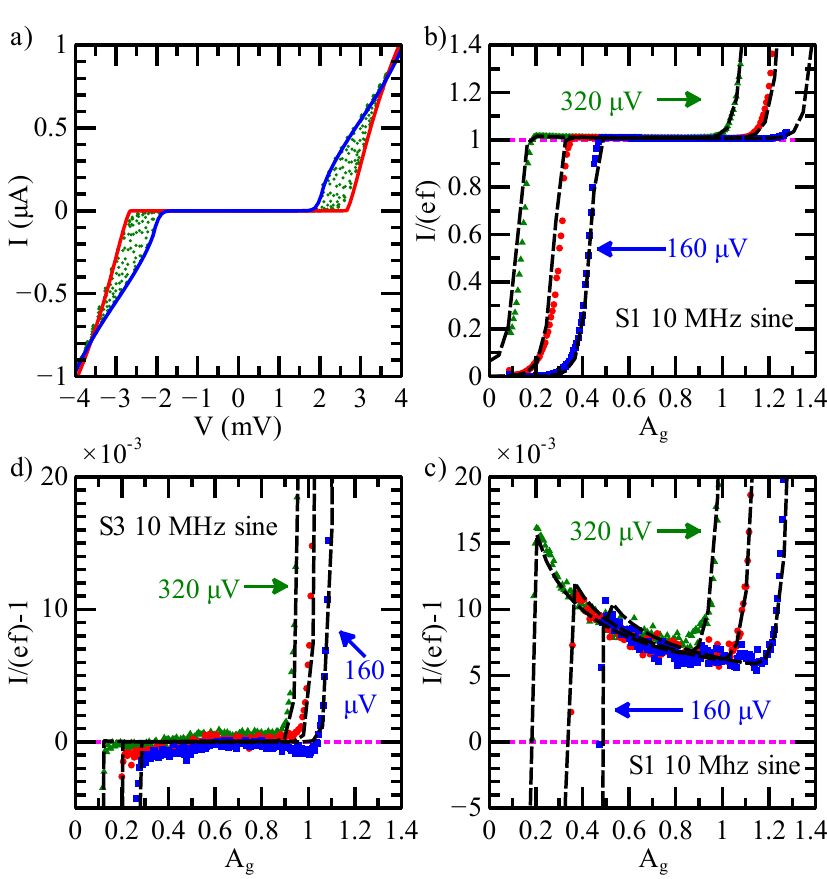}
\caption{\revision{10 MHz sine wave current pumping with simulations. (a) DC current vs voltage envelope measurement for sample S1 (green points). The red and blue lines are simulations of the measurement. (b) Current pumping plateaus for low charging energy sample S1 with a 10 MHz sine wave. (c) Close up of (b) showing the Andreev tunneling. (d) Close up of current pumping plateaus of sample S3 (high charging energy, low resistance) with a 10 MHz sine wave showing the absence of Andreev tunneling due to higher charging energy. 
Bias voltages are: 160 $\mu$V (blue), 240 $\mu$V (red) and 320 $\mu$V (green) with a magenta dotted line at $I=ef$ and simulations shown as black dashed lines in  figures b-d.}
}
\label{fig:pump10MHzsine2}
\end{figure}

\begin{figure}
\onefigure{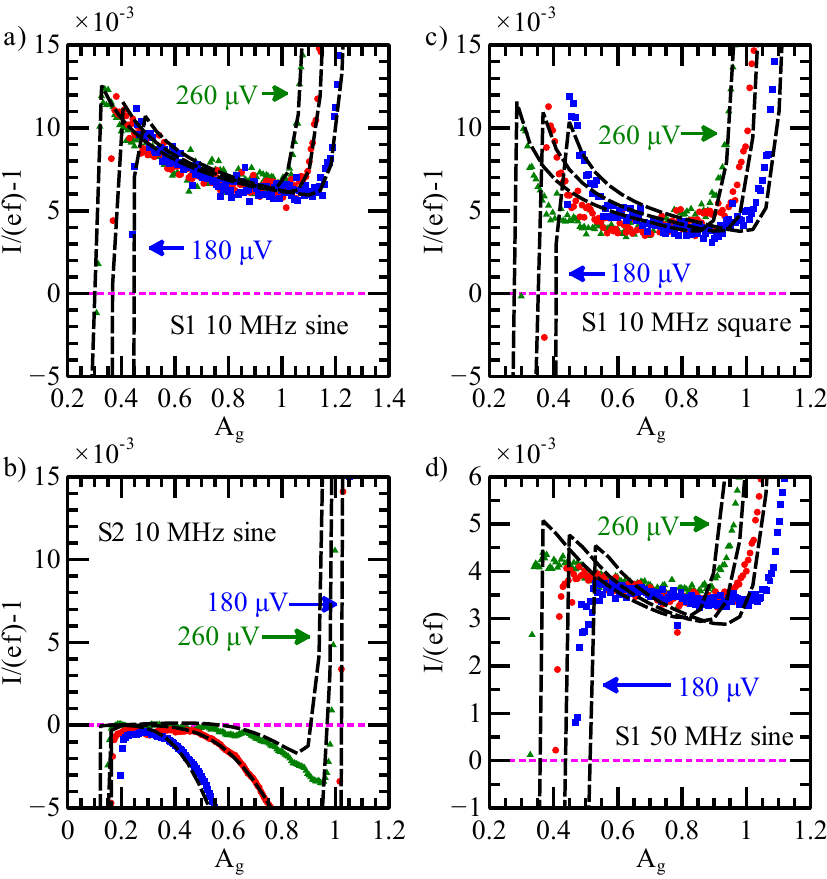}
\caption{\revision{Close ups of frequency and waveform dependence of current pumping plateaus (a) Pumping sample S1 with a 10 MHz sine wave showing the enhanced tunneling caused by the Andreev process. (b) Pumping of sample S2 (high charging energy, high resistance) with a 10 MHz sine wave showing the suppression of the enhanced Andreev tunneling. At high pumping amplitude, there is increased back tunneling compared to sample S3 due to the higher resistance of the junctions. (c) Pumping sample S1 with a 10 MHz square wave showing the different character of enhanced Andreev tunneling with a square wave drive.  (d) Close up of current pumping plateaus of sample S1 with a 50 MHz sine wave showing the smaller relative error at higher pumping frequency. The vertical scale has been changed compared to previous close-ups for clarity. The vertical scale in figures a-c are identical. The horizontal scale is not identical since the exact amplitude of RF drive applied to an individual device is not important. Bias voltages are: 180 $\mu$V (blue), 220 $\mu$V (red) and 260 $\mu$V (green) with a magenta dotted line at $I=ef$ and simulations shown as black dashed lines in all four figures. }
}
\label{fig:pump10MHzsine}
\end{figure}

\section{Experiment}
To form the SINIS turnstile, a normal copper island is connected to two superconducting aluminum electrodes using double angle evaporation with an intermediate oxidation step. Two types of devices were fabricated using electron beam lithography. The first type of sample used one step of lithography and aligned the device with a ground plane as shown in fig. \ref{fig:setup}(a). Large pads ensured capacitive coupling between the device and the ground plane, strongly suppressing environmental noise\cite{PekolaMaisi2010, Maisi2011}. These large pads are marked ground plane couplers (GPC) and are electrically isolated from the ground plane by atomic layer deposition (ALD) grown aluminum oxide. This device design allows simple single sample fabrication but is difficult to parallelize. A single gate line is used as both a DC gate and an RF gate.
 A scanning electron microscope (SEM) micrograph of the first type of sample is shown in fig. \ref{fig:setup}(b). 

 The second type of sample involves multiple steps of lithography as shown in fig. \ref{fig:setup}(c). In the first step, a ground plane is patterned to allow device fabrication on top of it. The turnstile sits in the patterned gap in the ground plane and only the leads are directly above it. Patterning the ground plane in this manner prevents formation of short circuits when wire bonding. After lithography, an insulating layer of silicon oxide is deposited by plasma enhanced chemical vapor deposition (PECVD). A second step of lithography is performed to form the turnstiles and individual DC gates for each turnstile.The lithography used a three layer mask with a hard germanium layer which allowed smaller features than the two layer PMMA patterning used for the first type.   The advantage of this fabrication method is that the separate DC gates enables parallelization. Parallelization is essential for metrological purposes as it is difficult to get large enough currents from a single device \cite{Maisi2009}. A SEM micrograph of the second type of sample is shown in fig. \ref{fig:setup}(d).

Measurements are done at a base temperature of approximately 70 mK. A typical current vs voltage envelope measurement reveals the superconducting gap and single electron transistor behavior of the turnstile \revision{as shown in fig. \ref{fig:pump10MHzsine2}(a)}. To measure the pumped current of the turnstile, a fixed bias voltage is applied to one of the turnstile's bias leads while current is measured on the other lead. A sine wave of amplitude $V_{AC}$ (resulting in a normalized amplitude $A_g=C_g V_{AC}/e$)  with a frequency $f$  is applied to the RF gate. 
As $A_g$ increases past a threshold value, the likelihood of transporting one electron through the turnstile in a single cycle approaches unity, giving rise to a quantized current plateau.

\revision{  We observe current quantization as shown in fig. \ref{fig:pump10MHzsine2}(b). Zooming in on the current plateau  reveals characteristic excess current above the expected $I=ef$ behavior due to Andreev tunneling. 
For samples with small charging energy ($E_C<\Delta$), as shown in fig. \ref{fig:pump10MHzsine2}(c), the current plateau is enhanced above the expected value particularly for low $A_g$ values in the plateau. These deviations are suppressed for samples with large charging energies ($E_C>\Delta$), as shown in fig. \ref{fig:pump10MHzsine2}(d). Amplifier gain is corrected on the $10^{-3}$ level to match the simulated plateaus. This does not influence the interpretation of excess current on the $10^{-2}$ level. Qualitatively, as $A_g$ is increased for the situation $E_C<\Delta$, we encounter an energy threshold that permits Andreev tunneling before the single electron tunneling threshold is encountered, allowing enhanced current flow for low charging energy samples. This charging energy dependence is indicative of the Andreev effect \cite{Saira2010, Maisi2011}.}

\revision{Fig. \ref{fig:pump10MHzsine} shows close ups of the current pumping plateaus for a different conditions. In fig. \ref{fig:pump10MHzsine}(a), pumping of sample S1 with a 10 MHz sine wave at a different range of biases than fig. \ref{fig:pump10MHzsine2} is shown. In fig. \ref{fig:pump10MHzsine}(b) the pumping curves for the high charging energy, high resistance sample S2 shows a reduction below the expected $I=ef$ due to single electron back tunneling.  This increased back tunneling at high pumping amplitude is a first order effect, which is due to the larger resistance of the high charging energy sample S2 compared to S3 \cite{Kemppinen2009apl}. Figs. \ref{fig:pump10MHzsine}(c) and (d) show sample S1 pumped with a 10 MHz square wave and a 50 MHz sine wave.}
 
 The Andreev effect is visible in a wide range of biases as \revision{shown in figs. \ref{fig:pump10MHzsine2}(b) and (c) and \ref{fig:pump10MHzsine}(a) with} no visible dependence on bias voltage.  Although the exact position of the plateau depends on bias, there is no separation of the plateaus due to bias dependence indicating this is not an environmental activation effect\cite{PekolaMaisi2010}. By fabricating high charging energy samples with low resistance, we can limit both the Andreev and back tunneling effects in the turnstiles as shown\revision{ in figs. \ref{fig:pump10MHzsine2}(d)}. This requires small junctions with highly transparent tunnel barriers i.e. junctions with low RC product as have been fabricated previously \cite{Brenning2004, Kemppinen2009apl}.
 
  \revision{It should be noted that for device S3 with optimized tunnel barriers for suppressing Andreev and single electron back tunneling there is still significant structure visible at the $10^{-3}$ level. The exact origin of this structure is not yet clear. It is possibly due to quasi-particle relaxation in the aluminum leads or noise still penetrating the filtering system. Preliminary simulations indicate that this structure can be accounted for if we consider overheated superconducting leads but future research is needed in this area. The focus here was to demonstrate control over the Andreev error process. SINIS pumps are theoretically predicted to achieve an accuracy of 1 part in $10^{8}$ with a current of 30 pA for a single pump \cite{Averin2008}. The experimentally measured accuracy level to date is on the $10^{-3}$ level i.e. 1 part in $10^3$.}

\section{Theory}
For the simulations shown in figs. \ref{fig:pump10MHzsine2} and \ref{fig:pump10MHzsine}, we calculate the average current flowing through the turnstile by numerically solving a master equation.
 The current through the left barrier (which is equal to the current through the right barrier in the steady state) is given by:
\begin{equation}
I=e \sum_n \left[ \Gamma_{LI}(n)-\Gamma_{IL}(n) \right] P(n,t)  
\label{eq:current}
\end{equation}
where $P(n,t)$ is the probability for finding $n$ electrons on the central island at time $t$. The tunneling rate from the left lead to the island is $\Gamma_{LI}(n)=\Gamma_{SN}(E_L^+(n))+2\Gamma_{AR}(E_L^{++}(n))$ and the tunneling rate from the island to the left lead is $\Gamma_{IL}(n)=\Gamma_{NS}(E_L^-(n))+2\Gamma_{AR}(E_L^{--}(n))$. The factor of two in front of the Andreev rate $\Gamma_{AR}$, compared to the single electron rates $\Gamma_{NS,SN}$, is to account for the Andreev process transporting two electrons. 
The energy required to add or remove a single electron to/from the island (denoted by $+$ or $-$) from the left or right lead (denoted by $L$ or $R$) is given by $E_{L,R}^{\pm}=\pm 2 E_C \left(n-n_g  \pm 1/2 \right) \pm eV_{L,R}$ where $E_C=e^2/2C$ is the charging energy with $C$ being the total capacitance of the island, $n$ is the number of excess electrons on the island and $n_g=C_g V_g /e$ is the normalized offset charge (the effective charge induced by applying a voltage $V_g$ to the gate with capacitance $C_g$) \cite{Averin1986}. The energy required to add or remove two electrons to/from the island (denoted by $++$ or $--$) in the Andreev process is given by $E_{L,R}^{\pm \pm}=\pm 4 E_C \left(n-n_g  \pm 1 \right) \pm 2 eV_{L,R}$.

\begin{table}
\caption{Sample Parameters.}
\label{table:paramtable}
\begin{center}
\begin{tabular}{cccccc}
Name  & $E_C/\Delta$ & $R_T$ (k$\Omega$) & $\Delta$ ($\mu$eV) & $A_{CH}$ (nm$^2$) \\
S1         &         0.63        &         160                &            220                    &        30        \\
S2         &         2.2          &         1400                &            210                    &      N/A      \\ %
S3          &         1.4        &         430                &            210                   &          N/A       \\ %
S4        &           0.75      &           110               &            210                   &          30        \\         

\end{tabular}
\end{center}
\end{table}

\begin{figure}
\onefigure{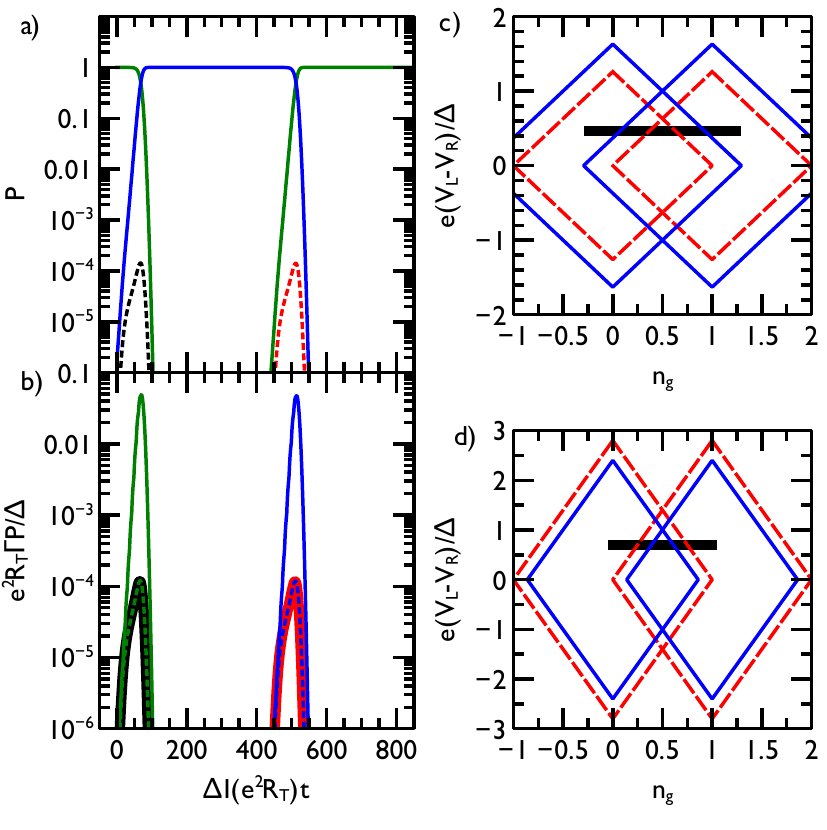}
\caption{\revision{(a) Probability evolution during charge pumping. The probability $P(n, t)$ is shown for different values of $n$: $n=0$ (green), $n=1$ (blue), $n=2$ (dotted black) and  $n=-1$ (dotted red).  (b)  Rates of tunneling weighted by probabilities of occupation. The single electron tunneling rate through the left junction from $n=0$ to $n=1$ is shown in solid green. The Andreev tunneling rate from $n=0$ to $n=2$ is shown in dotted green and is on top of the relaxation rate from $n=2$ to $n=1$ via single electron tunneling through the right electrode (thick black). Likewise, the single electron tunneling rate through the right junction from $n=1$ to $n=0$ is shown in solid blue, the Andreev rate $n=1$ to $n=-1$  is shown in dotted blue and is on top of the relaxation rate from $n=-1$ to $n=0$ through the left junction (thick red). (c) Stability diagram of sample S1. The blue diamonds are the thresholds for single electron tunneling, the red diamonds are the thresholds for Andreev tunneling and the thick black line is the pumping cycle used in (a) and (b).  (d) Stability diagram of sample S2. The diamonds are reversed in order because of the higher charging energy of sample S2.} 
    }
\label{fig:probs}
\end{figure}

The time dependence of the probability is given by the master equation\cite{Likharev1985, Averin1986}:
\begin{equation}
\begin{gathered}
\dfrac{dP(n,t)}{dt}=-\Gamma_{n,n}P(n,t) 
 \\ +\Gamma_{n-1,n}P(n-1, t) +\Gamma_{n+1,n}P(n+1,t) \\
+\Gamma_{n-2, n} P(n-2,t)+\Gamma_{n+2, n}P(n+2, t)
\label{eq:masterequation}
\end{gathered}
\end{equation}
where
\begin{align}
\Gamma_{n-1, n} &=\Gamma_{SN}(E_L^+(n-1))+\Gamma_{SN}(E_R^+(n-1)) \notag \\
\Gamma_{n+1, n} &=\Gamma_{NS}(E_L^-(n+1))+\Gamma_{NS}(E_R^-(n+1)) \notag \\
\Gamma_{n-2, n} &=\Gamma_{AR}(E_L^{++}(n-2))+\Gamma_{AR}(E_R^{++}(n-2))  \\
\Gamma_{n+2, n} &=\Gamma_{AR}(E_L^{--}(n+2))+\Gamma_{AR}(E_R^{--}(n+2)) \notag \\
\Gamma_{n, n} &=\Gamma_{n,n-1}+\Gamma_{n,n+1}+\Gamma_{n, n-2}+\Gamma_{n, n+2}.\notag 
\label{eq:gammas}
\end{align}

Equation \ref{eq:masterequation} tracks the flow of the probability. The higher order Andreev tunneling effects are included by considering state changes from $n$ to $n \pm 2$ \cite{Averin2008}.
The rate at which single electrons tunnel from superconductor to normal \revision{metal} $\Gamma_{SN}$, or the rate at which single electrons tunnel from the normal metal to the \revision{superconductor} $\Gamma_{NS}$,
are obtained using first order perturbation theory \cite{Ingold1992, Kemppinen2009}. 

The Andreev tunneling rates \revision{$\Gamma_{AR}$,} are given in equation 3 in reference \cite{Averin2008} using second order perturbative calculations. These rates depend on the charging energy $E_C$, the superconducting gap $\Delta$,  and the tunneling resistance $R_T$. Values for these parameters are obtained from DC current vs voltage envelope measurements (see inset of fig. \ref{fig:setup}) which depend only on $\Gamma_{NS,SN}$. $\Gamma_{AR}$ has an additional fitting parameter since its magnitude is controlled by the quantity $(\hbar/R_Te^2)/N$ where $N$ is the effective number of conduction \revision{channels,} $N=A/A_{CH}$\cite{Averin2008}. $A$ is the cross-sectional area of the junction estimated by scanning electron microscope (SEM) imaging (see fig. \ref{fig:setup} for relevant images and area estimates) and $A_{CH}$ is the effective area of a conduction channel used as a fitting parameter. Theoretically, $A_{CH} \approx 2$ nm$^2$ though fitted values are typically much larger than this value and are interpreted as resulting from inhomogeneities in the thickness of the oxide in the junctions\cite{Maisi2011}. In table \ref{table:paramtable}, the fitting parameters $E_C$ (in units of $\Delta$), $R_T$, $\Delta$ and $A_{CH}$ are listed for each of the samples simulated. Note that Andreev rate fitting parameters can not be determined for high charging energy \revision{samples S2 and S3} since the Andreev effect is suppressed below the measurement noise.

The simulated probability evolution during charge pumping is shown in fig. \ref{fig:probs}(a) for various charge states. The corresponding rates of tunneling weighted by the occupation probabilities are shown in fig. \ref{fig:probs}(b). These plots are for simulations of sample S1 with $V_L-V_R=200$ $\mu$V, $A_g=0.74$ and $f=10$ MHz. Only forward tunneling is relevant here. In figs. \ref{fig:probs}(c) and (d), we show the calculated stability diamonds for samples S1 and S2 respectively.  The minimal pair breaking energy for $1e$-tunneling is $\Delta$ so the threshold for tunneling is $E^{\pm}_{L,R}=\Delta$ which is shown as the blue line. For the Andreev tunneling, this pair breaking is avoided so the threshold for tunneling is $E^{\pm \pm}_{L,R}=0$ and is shown by the red dashed line.

\section{Results and Discussion}

In figs. \ref{fig:probs}(a) and (b), the probabilities and tunneling rates during one pumping cycle are shown for sample S1. As the transition rate from $n=0$ to $n=1$ grows, the probability for being in state $n=0$ quickly drops in the beginning of the cycle and is replaced with a probability for being in state $n=1$. At the same time, there is a detectable Andreev tunneling rate for going from state $n=0$ to $n=2$ after which the $n=2$ state quickly relaxes to $n=1$ by \revision{single-electron} tunneling. The low level occupation of the state $n=2$ is also detectable. Likewise, the opposite process where an electron leaves the turnstile is observable in the second part of the pumping cycle. As before, when the rate from $n=1$ to $n=0$ grows, the most likely state becomes $n=0$. The transition for $n=1$ to $n=-1$ by Andreev tunneling is also seen whereafter the $n=-1$ state quickly relaxes to $n=0$ state by \revision{single-electron} tunneling.

\revision{In figs. \ref{fig:pump10MHzsine2}(b), \ref{fig:pump10MHzsine2}(c) and \ref{fig:pump10MHzsine}(a), the pumping plateaus for six different biases and the corresponding fits are shown for low charging energy ($E_C<\Delta$) sample S1. There is excellent agreement between the fits and the data. By looking at the stability diagram shown in fig. \ref{fig:probs}(c), we can qualitatively understand the observed effect. The stability diagram shows tunneling thresholds versus normalized bias voltage, $V$, and momentary gate charge, $n_g$.  Starting in the diamond on the left, as the gate amplitude, $A_g$, is gradually increased, the Andreev threshold (shown as a red dotted line) is encountered first. Thus the Andreev process is most noticeable at low $A_g$. As $A_g$ is increased further, we encounter the single electron tunneling threshold and this process quickly dominates, obscuring the Andreev effect. This results in the enhanced current pumping plateau in figs. \ref{fig:pump10MHzsine2}(c) and \ref{fig:pump10MHzsine}(a). For comparison, as can be seen in fig. \ref{fig:probs}(d), the high charging energy ($E_C>\Delta$) sample, S2, encounters the single electron tunneling threshold before the Andreev threshold. The Andreev effect is not observed as the electron has already tunneled before it enters the Andreev regime. Thus the plateau in figs. \ref{fig:pump10MHzsine}(d) and \ref{fig:pump10MHzsine2}(d) is flat with no evidence of the enhanced tunneling effect. }

 The lack of dependence on bias seen in \revision{figs. \ref{fig:pump10MHzsine2}(c) and \ref{fig:pump10MHzsine}(a) can similarly} be explained by looking at the stability diagram. Changing the bias corresponds to changing the horizontal level of the \revision{black} line in fig. \ref{fig:probs}(c). However, the Andreev tunneling thresholds run parallel to the single electron tunneling thresholds so this has a very small effect on the observed current.

\revision{In fig. \ref{fig:pump10MHzsine}(c), }the results from pumping with a 10 MHz square wave \revision{for sample S1} are shown. The square wave is modeled with an exponential rise to the applied voltage with risetime 2.5 ns. Compared to the sine wave pumping of the same frequency, the enhanced current for the square wave is more peaked and dies away more quickly. \revision{At an amplitude of $A_g=0.75$, the square wave pumped plateau is almost flat with a current less than $1.005ef$ (i.e. closer to the ideal value of $I=ef$) while sine wave pumping with the same frequency and bias voltage has a plateau with significant slope at $A_g=0.75$ with a current greater than $1.005ef$}. This behavior can also be qualitatively understood from the stability diagram. The square wave goes more immediately to the final $A_g$ value thus spending less time in the vulnerable regions of the stability diagram. When we first encounter the Andreev threshold but before the single electron threshold is encountered, the Andreev process dominates and we see the sharp peak in current shown \revision{in fig. \ref{fig:pump10MHzsine}(c).} As $A_g$ is increased, we encounter the \revision{single-electron} tunneling process threshold and the escape process is dominated by it. There is only a small possibility for the Andreev tunneling enhancement to occur since the square wave sweeps through that vulnerable region quickly. Thus the Andreev effect in the pumping plateau dies away more quickly for the square wave than for the sine wave.

\revision{In figs. \ref{fig:pump10MHzsine}(d), }the pumping plateaus for a 50 MHz sine wave are shown with the Andreev effect visible for low charging energy sample, S1. With higher frequency, more electrons are pumped per unit time producing a higher current. Thus the relative accuracy improves compared to the lower frequency 10 MHz sine wave shown in \revision{fig. \ref{fig:pump10MHzsine}(a)} since the absolute value of the excess Andreev current remains roughly the same.  The Andreev tunneling contribution to the current remains roughly unchanged because the same process as described for the lower frequency sine wave takes place. \revision{This results in the more flat slope of the 50 MHz pumping seen in figure \ref{fig:pump10MHzsine}(d) but the Andreev process is still detectable}. It should be noted that the same sample parameters listed in table \ref{table:paramtable} were used for simulating 10 MHz sine wave, 10 MHz square wave and 50 MHz sine wave current pumping demonstrating the robustness of the simulation. 

From the simulations, we extract an effective channel area on the order of 30 nm$^2$ for the pumping (the effective channel area is difficult to determine precisely so the values are rounded to the nearest 10's of nm$^2$). \revision{This was consistent for both samples S1 and S4 shown in figure \ref{fig:setup}. For clarity of presentation, only the data and fits from sample S1 are shown in figures \ref{fig:pump10MHzsine2} and \ref{fig:pump10MHzsine}. Sample S2 and S3 do not have a discernible Andreev parameter as it is suppressed by their high $E_C$.} This is in general agreement with the effective area of 30 nm$^2$ found in earlier work\cite{Maisi2011}. These values are approximately one order of magnitude larger than the theoretical channel area of approximately \revision{2 nm$^2$,} indicating that roughly only one tenth of the effective area of the junction is active in agreement with previous results \cite{Greibe2011, Pothier1994}.

We have shown that the enhanced current is dependent on $E_C$, pumping amplitude and pumping waveform shape but independent of pumping frequency and bias voltage. These are all characteristic signatures of Andreev tunneling that can be well accounted for by our theoretical model.

\section{Conclusion}
We have observed the Andreev tunneling process in current pumping with a single electron turnstile. This error process can be effectively suppressed with high charging energy leading us one step closer to a quantum current standard and completing the quantum metrological triangle. \revision{Andreev reflection as an error process can be fully eliminated with proper choice of $E_C$ as demonstrated by these experiments.}

\acknowledgments
This work was funded in part by the European Community's Seventh Framework Programme under Grant Agreement No. 218783 (SCOPE), the Aalto University Postdoctoral Researcher Program and the Finnish National Graduate School in Nanoscience.
\bibliographystyle{eplbib}

\end{document}